# Percentage Coefficient ($b_p$)

# -- Effect Size Analysis (Theory Paper 1)

**Authors:** Xinshu Zhao[1*]; Dianshi Moses Li[2]; Ze Zack Lai[1]; Piper Liping Liu[3]; Song Harris Ao[1]; Fei You[1]

**Affiliations:**

[1] Department of Communication; Faculty of Social Science; University of Macau, Macao

[2] Centre for Empirical Legal Studies; Faculty of Law, University of Macau, Macao

[3] School of Media and Communication, Shenzhen University, China

[*] Corresponding author. Email: xszhao@um.edu.mo

**Percentage Coefficient ($b_p$)**

**-- Effect Size Analysis (Theory Paper 1)**

## Abstract

Percentage coefficient ($b_p$) has emerged in recent publications as an additional and alternative estimator of effect size for regression analysis. This paper retraces the theory behind the estimator. It's posited that an estimator must first serve the fundamental function of enabling researchers and readers to comprehend an estimand, the target of estimation. It may then serve the instrumental function of enabling researchers and readers to compare two or more estimands. Defined as the regression coefficient when dependent variable (DV) and independent variable (IV) are both on conceptual 0-1 percentage scales, percentage coefficients ($b_p$) feature 1) clearly comprehendible interpretation and 2) equitable scales for comparison. The coefficient ($b_p$) serves the two functions effectively and efficiently. It thus serves needs unserved by other indicators, such as raw coefficient ($b_w$) and standardized beta ($\beta$).

Another premise of the functionalist theory is that "effect" is not a monolithic concept. Rather, it is a collection of concepts, each of which measures a component of the conglomerate called "effect," thereby serving a subfunction. Regression coefficient ($b$), for example, indicates the unit change in DV associated with a one-unit increase in IV, thereby measuring one aspect called *unit effect*, aka *efficiency*. Percentage coefficient ($b_p$) indicates the percentage change in DV associated with a whole scale increase in IV. It is not meant to be an all-encompassing indicator of an all-encompassing concept, but rather a comprehendible and comparable indicator of *efficiency,* a key aspect of *effect*.

**Percentage Coefficient ($b_p$)**

**-- Effect Size Analysis (Theory Paper 1)**

The world is not black or white. It has shades of grey. It is continuous, colorful, and complex. The human brain, however, is too limited to process the world as it is. To cope with this fundamental mismatch, we simplify, and we dichotomize. We see black or white. We think black or white. We talk, hear, write, and read black or white. We make tools – concepts, symbols, languages, theories, mathematical formulars, statistical models, etc., -- to enable and ease the simplification. But we also overuse, misuse, and abuse the tools. We oversimplify, and we over-dichotomize. We interact with the tools and with each other to recycle and reinforce the black-or-white view of the world. We produce, propel, and perpetuate the selective spirals of over-simplification, knowingly or not.

*P*-values and *null hypothesis significance testing* (NHST) is one such tool. As probably the most used tool in scientific enquiry of modern time, it simplifies the concept of effect, turning it into two dichotomies – 1) sufficient vs. insufficient evidence for effect, and 2) positive vs. negative effect. But the real-world effects are not dichotomous. They are continuous, having varying degrees of strengths and, as we hope to show with this series of studies, they have various components and compartments. To many, the mismatch between the simplified and the real worlds is a root cause of the so-called statistical crises, aka *p*-value crises.

An array of remedies has been proposed. One that has been broadly accepted is to report effect sizes and confidence intervals, which is to replace or supplement *p* values and NHST. It made sense. While *p*-values and NHST over-dichotomize the numerical world, reporting numerical sizes should re-present the real world more like it is, alleviating the mismatch.

However, it didn't happen as expected or hoped. While researchers of health behavior followed the instructions to report regression coefficients and confidence intervals, few interpreted or compared them to aid theory building or practical suggestion. The coefficients occupied much print space. Otherwise, they did not do much. Although the researchers performed their duty of reporting the effect size indicators, the indicators failed to perform their duty of depicting the numerical world.

Now, nearly a quarter century after the APA Task Force report kickstarted the effect size movement, the movement seems stalled (Wilkinson & APA_Task_Force, 1999). The overreliance on *p*-values and NHST continues (Collaboration, 2015; Head et al., 2015; Jasny et al., 2017; Masicampo & Lalande, 2012; Munafò et al., 2017; Wooditch et al., 2020).

This article introduces a functionalist theory to explain what's missing in the current application of regression coefficients. The theory guided our design of *percentage coefficient* ($b_p$), a reconfigured regression coefficient, to fill some void. A real-data example illustrates its application.

We plan this study as the first of a series of studies on effect size indicators and their applications in analyzing data and better understanding the health behavior. The studies will also discuss or display the roles of *p*-values and NHST in effect size analysis.

## I. *P*-value, Significance Testing, and Effect Sizes

Statisticians and scientists were aware of the limitations of null-hypothesis significance testing (NHST) and *p*-values (Baker, 2016; Berkson, 1938; Chin et al.,

2023; Cohen, 1994; Loftus, 1991; Loftus, 1996; Lykken, 1968; Makel et al., 2012; Meehl, 1954; Meehl, 1992; Munafò et al., 2017; Simmons et al., 2011; Snyder & Lawson, 1993; Thompson, 2004; Wooditch et al., 2020). A root cause of the problem is the binary world view that oversimplifies the continuous world mismatches the mission of quantitative science (Hayes, 2022; Liu & Li, 2023). To compensate and alleviate, professional organizations and academic leaders recommend reporting effect sizes and confidence intervals as replacement or supplement to the *p*-values and NHST (AERA_Task_Force, 2006; Cohen, 1990, 1994; Kirk, 1996; Preacher & Kelley, 2011; Thompson, 1996; Thompson, 1998, 1999; Wilkinson & APA_Task_Force, 1999; Zientek et al., 2016).

A prominent task force in the psychology research community made the following recommendation:

> Always present effect sizes for primary outcomes...If the units of measurement are meaningful on a practical level (e.g., number of cigarettes smoked per day), then we usually prefer an unstandardized measure (regression coefficient or mean difference) to a standardized measure (r or d).

By one count at the beginning of the century, in just education and psychology, 17 journals announced or amended editorial policies requiring effect size information (Robinson et al., 2002).

Researchers responded by reporting regression coefficients, mostly standardized beta ($\beta$) or unstandardized raw $b$ ($b_w$), and their confidence intervals. An *effect size movement* ensued (Kelley & Preacher, 2012; Robinson et al., 2002; Robinson et al., 2003). Nevertheless, there have been few advice or guidelines on how to interpret the regression coefficients, or how to use them to address research questions, build theories, or guide decisions. In published studies, the numerical values of the coefficients were rarely interpreted or used other than filling up the main cells of the tables. The directional signs, + or -, were the only parts regularly used, and they were used to help testing the binary hypotheses. The dominance of NHST and *p*-values continues, and so does the binary world view.

It's no surprise, therefore, that the criticisms of NHST and *p*-values has grown even more intense more recently (Amrhein et al., 2019; Benjamini et al., 2021; Halsey et al., 2015; He et al., 2019; Huang, 2023; Matthews, 2021; Matthews et al., 2017; Nuzzo, 2014; Wasserstein & Lazar, 2016; Wasserstein et al., 2019), highlighted by one journal banning *p*-values and NHST (Trafimow, 2018; Trafimow & Marks, 2015; Woolston, 2015).

This study proposes a theory that posits two primary functions of statistical indicators, 1) to aid comprehension and 2) to aid comparison. The functionalist theory explains why regression coefficients in their current forms ($\beta$ and raw $b$) are incapable of performing the functions, therefore incapable of accomplishing the missions accorded to them. The theory also defines the qualities that are needed in a renovated coefficient if it is to fill some of the void.

The article then introduce *percentage coefficient* ($b_p$) a product of integrating elements of three traditions 1) functionalist view of statistical indicators inspired by social science theories, 2) relational view of meanings and symbols influenced by semiotics philosophy and 3) scale normalization techniques from data mining and machine learning research under the umbrella of computer science and data science.

We will present a real-data example to demonstrate the application of $b_p$, and to compare its performance with that of $\beta$ and raw $b$. We will show that $b_p$ can effectively aid comprehension and comparison; and can provide information otherwise unavailable.

We will discuss the assumptions and limitations of $b_p$. With its assumptions understood and limitations addressed, we will speculate that $b_p$ may help to revive and re-energize the effect-size movement, thereby helping to put NHST and *p*-values back to their proper roles.

## I. Two Functions of Regression Coefficients – Comprehension and Comparison

We begin by recognizing two primary functions of regression coefficients (*b*), comprehension and comparison. The *comprehension* function requires that a regression coefficient enables researchers and readers to interpret and comprehend the effect size that the efficient represents. The *comparison* function requires that researchers and readers may contrast and compare two or more regression coefficients to evaluate the relative sizes of the effects that the coefficients represent. Each function requires a certain scale unit, which we discuss below.

### I.1. The Concept of *Unit*

To understand the functions of regression coefficient we need to first explicate “scale unit” A *scale unit* is a segment in a scale that is given the numerical value 1. For example, variable *age* measured by year has “year” as the scale unit. Variable *annual income* measured by dollars has “dollar” as the unit, while the same variable measured by thousand dollars has “thousand dollars” as the unit.

We can discuss these examples efficiently because the units in each example have a uniform meaning. Of the age scale, for instance, the difference between 18 and 19 is one *year* just like the difference between 99 and 100. The same is true for all other units. Not all scales have the same feature. The segment between “strongly agree” and “agree” in a five-point Likert scale, for example, does not have the same meaning as the segment between “agree” and “neutral.” Such scales lack uniformly meaningful units.

Regression coefficients based on such scales are not capable of carrying out their duties, as we discuss below.

### I.2. Comprehension Function and Uniform-Meanings Assumption

There are two primary missions of a statistical indicator. The first is to enable and ease comprehension. This mission requires interpretability, i.e., the meaning of the indicator being clear to researchers, reviewers, and readers. A regression coefficient (*b*), defined as the unit change in the dependent variable (DV) associated with a one-unit increase in the independent variable (IV), can be interpretable only if the underlying dependent (DV) and independent (IV) scales each has a *uniformly meaningful unit*. That means the meaning of the unit is clear and stays consistent between segments of the scale, which we will refer to as *uniform-meanings* assumption. The plural, meanings, imply two points.

1) Meaning has two components, verbal and numerical. Each needs to be clear.
2) There are two scales, for DV and IV, therefore two sets of units. Each set needs a uniform meaning.

Suppose *b*=50 for age (year) predicting healthcare expenditure (dollar). It indicates a $50 increase in expenditure associated with a one-year increase in age. This *b* coefficient is interpretable because each unit in the DV scale has a uniform meaning, *dollar*, and each unit in the IV scale has a uniform meaning, *year*, satisfying the requirement for *uniform meaning*. The two uniform meanings make *b* interpretable, enabling it to perform the function of enabling comprehension.

*Uninterpretable Raw b ($b_w$).* A raw coefficient (raw *b*, $b_w$) is a regression coefficient based on under-processed DV or IV scales – the scales inadequately processed for the purpose of comprehension and/or comparison. Raw *b* ($b_w$) is typically uninterpretable because the dependent scale or the independent scale lacks a clear and uniform meaning, violating the requirement for uniformed meaning of units (UMU). Suppose $b$=0.2 for social media use (1-4 ordinal, never-sometimes-often-daily) predicting feeling depressed (1-5 Likert, SD-disagree-neutral-agree-SA). The meaning of the units may differ between segments, e.g., between agree and neutral and between agree and strongly agree. There is no clear uniform meaning for either scale. Ordinal scales, including Likert scales, lack clear uniform meaning unless redefined, transformed, or otherwise processed to produce a uniform meaning.

*Practically Uninterpretable $\beta$.* Standardized beta ($\beta$) is a regression coefficient with *standard deviation* (SD) as the scale unit for DV and IV. It is therefore mathematically interpretable – it indicates the change in DV measured by SD associated with a one-SD increase in IV. Nevertheless, SD is *only* a mathematical concept. It was not designed to contain or convey a meaning of substance for theory building or decision making. Therefore, $\beta$ is generally not interpretable for research practice, which was one reason that statisticians from early on objected $\beta$ as a general indicator of effect size (Achen, 1982; Bring, 1994; Darlington, 1968, 1990; Greenland et al., 1991; Greenland et al., 1986; Hanushek & Jackson, 1978, 2013; Kim & Ferree Jr, 1981; King, 1986; Pedhazur & Kerlinger, 1982; Tukey, 1954). That may explain why publications rarely if ever interpret $\beta$ statistics even though they are routinely reported and often occupy the largest space of regression tables. That may also explain why the APA Task Force, while often credited as a significant force pushing for the "effect size movement," advised against standardized measures, including $\beta$, to indicate effect size (Wilkinson & APA_Task_Force, 1999).

*Need for comprehendible b*. The above analysis shows the need for a new coefficient (*b*) based on uniformly meaningful units, which should make *b* interpretable, i.e., comprehendible. Percentage coefficient ($b_p$), as we will show, may serve the need.

**I.3. Comparison Function and Equitable-Units Assumption**

The second function of statistical indicator is to enable cross-comparison between two or more objects represented by the indicators. Two regression coefficients can be compared with each other if and only if both parts of the *equitable-units* assumption hold true --

1) The DV units of the compared variables are equitable to each other.

2) The IV units of the compared variables are equitable to each other.

But the units must have uniform meanings before we can consider, i.e., to think and talk about, whether the units are equitable. In other words, uniform meanings are necessary conditions for equitable units.

Suppose we compare *b* = - 0.06 for hours/week of aerobic exercise with *b* = - 0.03 for hours/week of anaerobic exercise, both predicting weight loss (kg). Suppose we infer that aerobic exercise is twice as efficient as anaerobic exercise. The comparison assumes an hour of one type of exercise is equitable to an hour of another type of exercise (Condition 2 above), which may be a reasonable assumption depending on objectives and perspectives. Condition 1 above is easier to meet for this comparison because the comparison involves only one DV, body weight.

Suppose we use the number of smokers to predict cases of lung cancers and cases of rheumatoid arthritis. Coefficient *b*=0.9 for lung cancer vs *b*=0.1 for rheumatoid arthritis would suggest that smoking causes eight times more cancer than arthritis. The comparison equates a case of cancer with a case of arthritis. Depending on research objective and context, the equating may or may not be justified.

*Incomparable Raw b ($b_w$).* As said, unprocessed or under-processed raw scales often lack uniformly meaningful units. Units without uniform meanings cannot be considered equitable to each other. Even when two scales both have uniform meanings, e.g., education measured by year and income measured by dollar, the units are usually not equitable, e.g., one year of education cannot be equated with one dollar of income (Afifi & Clarke, 1990; Walsh, 1990) . This explains why publications rarely compare raw coefficients ($b_w$), the regression coefficients unprocessed or under-processed for comparison, even when the coefficients occupy the main space of regression tables.

*Incomparable $\beta$.* Standardized beta ($\beta$) has standard deviation (SD) as the scale unit, giving many the impression that SD of one variable ($SD_{v1}$) can be equated with SD of another variable ($SD_{v2}$). Unfortunately, $SD_{v1}=SD_{v2}$ is a statistical illusion, and the impression is wrong. As SD was not designed to contain meaning of substance, two SDs cannot be considered as equitable to each other for general practice or decision making. As a result, generally $\beta$ coefficients cannot be compared with each other when assessing effect sizes.

Suppose SD of aerobic exercise is one hour, and SD of anaerobic exercise is half of hour. A direct comparison of two $\beta$ coefficients involving the two independent variables would assume 1=0.5, which would be hard to justify despite the widespread beliefs in the equation $SD_{v1}=SD_{v2}$. Giving the illusion of being a comparable indicator of effect sizes has been a main reason that experts objected $\beta$ (Achen, 1982; Bring, 1994; Darlington, 1968, 1990; Greenland et al., 1991; Greenland et al., 1986; Hanushek & Jackson, 1978, 2013; Kim & Ferree Jr, 1981; King, 1986; Pedhazur & Kerlinger, 1982; Tukey, 1954).

Better known is that standard deviation changes over time. Suppose in a panel survey, SD=1.0 for hours of physical activity in Wave 1 and SD=1.1 for the same variable in Wave 2. Comparing *β* coefficients involving the two IV would mean 1 hour =1.1 hour, which would be harder to justify than even the equation $SD_{v1}=SD_{v2}$. In other words, *β* relies on a parameter (SD) that varies easily without real change in effect size. That is the main reason for considering *β* an invalid indicator of effect size (Achen, 1982; Bring, 1994; Darlington, 1968, 1990; Greenland et al., 1991; Greenland et al., 1986; Hanushek & Jackson, 1978, 2013; Kim & Ferree Jr, 1981; King, 1986; Pedhazur & Kerlinger, 1982; Tukey, 1954).

Published studies rarely if ever compare *β* coefficients to assess differences in effect sizes, even though they often occupy much table space. It was right not to compare *β* coefficients given the incomparability. However, it wastes scholarly and social resources to publish statistics that perform no real function other than filling table space and superficially meeting the requirement of publishing them.

*Need for comparable b*. The above analysis shows the need for a new coefficient (*b*) based on equitable units, which should make *b* coefficients comparable to each other. Percentage coefficient ($b_p$), as we will show, was designed to serve the need.

*Need for informative confidence intervals*. In addition to reporting effect sizes, researchers are urged or required to report confidence intervals. A confidence interval is a range of effect sizes. When the effect size indicators, $b_w$ and *β*, are uninterpretable and incomparable, their confidence intervals are uninterpretable and incomparable. When regression coefficients fail their duties, confidence intervals cannot perform theirs.

### I.4. Fixed Referencing and Cross Referencing

Both functions entail referencing. Comprehension requires *fixed referencing*. To understand a value is to contrast it with a fixed reference point, the scale unit. To understand the value of US $2,000 is to contrast it with the US $1, the scale unit. Travelers would not understand the value of 2,000 Seychellois Rupee if they do not know the value of one Seychellois Rupee. To comprehend the meaning of $b$=0.5 is to contrast it with $b$=1. However, $b$=1 could not be understood unless the meanings of DV and IV units are both understood. Inadequately processed raw $b$ ($b_w$) cannot be comprehended when the meanings of the fixed reference points, i.e., the scale units of DV and IV, are not understood, hence the meaning of another reference point, $b$=1, is not understood.

Comparison requires *cross referencing*. That means to contrast two or more values against each other, making them serve as each other's reference points. The two values are cross-comparable if and only if (iff) certain scale units are deemed equitable to each other.

Suppose one says one I-phone is more expensive than one Android phone because the I-phone costs $1,000 while the Android phone costs $800. He is equating $1 with $1, and one I-phone with one Android phone. Suppose he says the Android brand raised price because it was $780 five years ago. He is equating $1 now with $1 five years ago, ignoring inflation; he is also equating one phone today with one phone five years ago, ignoring upgrades.

Suppose one compares two $b$ coefficients, $b_1$ and $b_2$. He is assuming three equalities, 1) one $b_1$ equals one $b_2$, 2) the two DV scale units equal each other, and 3) the two IV scale units equal each other. By assuming the equalities, he is cross-referencing the scales.

### I.5. Un-interpretability and Incomparability Nullify Effect Size Movement

In response to the calls for reporting sizes, researchers routinely report raw *b* ($b_w$), standardized beta ($\beta$), and confidence intervals. But they rarely interpret or compare coefficients. The coefficients cannot be easily interpreted or compared. As a result, researchers continue to rely on the *p* values and NHST, which dichotomizes the continuous reality. The un-interpretability and incomparability of the effect-size indicators nullified the effect size movement, extending the pitfalls and perils of the *p* values and null-hypothesis significance testing (NHST).

### I.6. Section Summary

The mission of regression coefficients, like that of other numerical indicators, is to aid comprehension and comparison. To allow comprehension, a coefficient needs to be interpretable, which requires its scale units to have a uniform meaning. To allow comparison, the underlying scales need to have equitable units, which also requires a uniform meaning of the units. Raw *b* ($b_w$) and standardized beta ($\beta$) do not usually meet the two conditions – their scale units tend not to be uniformly meaningful, and they tend not to be equitable to each other. As a result, raw *b* ($b_w$) and standardized beta ($\beta$) tend to be uninterpretable and incomparable in research practices of some disciplines.

In short, $b_w$ and $\beta$, the two most often used regression coefficients, are not equipped to fulfill the missions of aiding comprehension and comparison. This manuscript discusses an emerging alternative, *percentage coefficient* ($b_p$), that has been designed and developed to accomplish the missions.

## II. Percentage Coefficient ($b_p$) for Comprehension and Comparison

A *percentage coefficient* ($b_p$), in the original form, is a regression coefficient (*b*) when the dependent variable (DV) and the independent variable (IV) are each on a 0~1 percentage scale. The extended percentage coefficient ($b_{p\pm}$) features extended percentage scales ranging -1 ~ +1. This study is focused on the basic form, $b_p$. Thus, to understand percentage coefficient one needs to first understand percentage scale, which we turn to next.

### II.1. Defining Percentage Scale

A *percentage scale* is a scale anchored by numbers 0 and 1 where 0 represents conceptual minimum and 1 represents conceptual maximum. A *conceptual anchor* may differ from the corresponding *observed anchor*, i.e., a conceptual maximum may differ from the observed maximum, and a conceptual minimum may differ from the observed minimum. A percentage scale for age may have 0 for newborns, and 1 for 100 years old; but the observed minimum may be 18 years old, which would be 0.18 on percentage scale, and the observed maximum may be 70 years old, namely 0.70 $s_p$. Conceptual maximum or conceptual minimum may also differ from *population maximum* or *population minimum*. Extending the example, the oldest person in the population may be 110 years old, making 1.1 $s_p$ the population maximum.

### II.2. Development of Percentage Scale and Percentage Coefficient ($b_p$)

Scales with 0~100 conceptual range can be traced several centuries back. In 1701, English polymath Isaac Newton (1642-1726) devised thermometer to measure heat. The Newton scale was anchored at the lower end by 0, which represents the temperature for water freezing. Over 20 years later, in 1724, Daniel Gabriel Fahrenheit (1686 – 1736), a Dutch physicist born in Poland to a family of German extraction, constructed his thermometer that also has 0 as the lower anchor. Now known as the Fahrenheit scale, it ranged conceptually 0~90 °F, where 0 °F was the freezing temperature of brine mixture made of water and ammonium chloride salt, and 90 °F was the average body temperature of humans. ("Fahrenheit Gabriel Daniel," 1911) ("Fahrenheit temperature scale," 2023). In 1742, Swedish astronomer Anders Celsius (1701–1744) developed a 0~100 scale where 0 °C was the freezing temperature of water and 100 °C was the boiling temperature of water. ("Celsius temperature scale," 2023) Now the official temperature scale of almost all countries in the world today, the Celsius scale is not only one of the oldest percent scales (0~100); it is also one of the most impactful.

In the 1980s and 1990s, researchers of media and advertising effects used Equation 1 to linearly transform their DV, IV or both to 0~100 percentage scales to interpret or compare the regression coefficients (Thorson et al., 1987; Thorson & Zhao, 1997; Zhao, 1989, 1997; Zhao & Xie, 1992; Zhao et al., 1994).

$$s_p = \frac{o_s - m_n}{m_x - m_n} 100 \qquad \text{Eq. 1}$$

$s_p$: percentage score after scale transformation
$o_s$: original score before scale transformation
$m_x$: measurement maximum on original scale
$m_n$: measurement minimum on original scale

Later, to compare continuous IVs with dichotomous IVs on 0-or-1 dummy scales, the 0~100 scales were switched to 0~1 scales using a formular equivalent to Eq. 2 (Zhao et al., 2010; Zhao & Zhang, 2014). Zhao and Zhang (2014) recommended to transform all variables – DV, IV and control variables, numerical and dummy variables – to 0-1 scales. They coined the term *percentage scale* to represent 0-1 scales, including 0-or-1 dummy scales and 0~1 continuous scales, and the term *percentage coefficient* ($b_p$) to represent the regression coefficient on such scales.

$$s_p = \frac{s_o - c_n}{c_x - c_n} \qquad \text{Eq. 2}$$

$s_p$: percentage score after scale transformation
$s_o$: original score before scale transformation
$c_x$: conceptual maximum on original scale
$c_n$: conceptual minimum on original scale

In the meantime, feature scaling and scale normalization techniques advanced in the fast-advancing field of data mining and machine learning (Han et al., 2012; Han & Kamber, 2001; Jain et al., 2005). Of the normalization methods, the min-max normalization used Eq. 3 for linear transformation (Shalabi et al., 2006).

$$v' = [\frac{v - min_o}{max_o - min_o}\ (max_n - min_n)] + min_n \quad (min_n \leq v' \leq max_n) \qquad \text{Eq. 3}$$

$v'$: new score after scale transformation
$v$: original score before scale transformation
$max_o$: maximum on original scale
$min_o$: minimum on original scale
$max_n$: maximum on new scale
$min_n$: minimum on new scale

Percent scale (0~100, Eq. 1) and percentage scale (0~1, Eq. 2) are special cases of min-max scales (Eq. 3). This is shown in that 1) Eq. 3 becomes Eq. 1 when $min_n$=0 and $max_n$=100 and 2) Eq. 3 becomes Eq. 2 when $min_n$=0 and $max_n$=1. This mathematical relationship reflects the scales' functions. Min-max scales were developed as preprocessing tools to prepare scales before more focused operations (Patro & Sahu, 2015; Shalabi et al., 2006). Percentage scales, by contrast, were designed as the core of an operation focused on making percentage coefficients ($b_p$) comprehensible and comparable.

The two lines of research, however, share the need to equalize scales – feature scaling needs it to equalize feature weights when teaching machines, while percentage scaling needs it to level the ground for comparison. The success of feature scaling in machine learning provides a powerful validation that percentage scaling (Eq. 2) indeed equalizes scales. It affirms the equitable-unit assumption (I.3. above), a key necessary condition for comparing statistical indicators such as percentage coefficients ($b_p$). Encouraged, more studies reported percentage coefficients ($b_p$) as a main indicator of effect sizes (Ao et al., 2023; Han et al., 2023; Jiang et al., 2021; Liu, Chang, et al., 2023; Liu, Ye, et al., 2023; Zhang, Ao, et al., 2023; Zhang, Ao, & Zhao, 2023; Zhao et al., 2023; Zhao, Ye, et al., 2022).

## II.3. Defining Percentage Scale and Percentage Coefficient ($b_p$)

*Percentage scale.* As per above discussion, we define *percentage scale* as a scale anchored by a conceptual maximum coded 1 ($c_x$=1) and conceptual minimum coded 0 ($c_n$=0), which may be reconstructed from another scale using Eq. 2.

*Percentage coefficient* ($b_p$). We define *percentage coefficient* ($b_p$) as a regression coefficient when both dependent and independent variables are on percentage scales.

## II.3. Percentage coefficient ($b_p$) is interpretable, enabling comprehension.

As discussed, the necessary and sufficient condition for interpretability of a regression coefficient is that its scale unit be uniformly meaningful. The unit of a percentage scale has a clear meaning: it is the whole scale, i.e., one hundred percent (100%) of the scale. The entire scale has just one unit. The meaning is uniform by the conception and construction of the scale. As it is based on DV and IV scales with uniformly meaningful units, percentage coefficient ($b_p$) is always interpretable. Over the years researchers have developed various expressions that all represent the same mathematical entity. The following are four variations.

**Interpretation 1: Effect per unit.** Coefficient $b_p$ represents *effect per unit*, aka *unit effect*, where *unit* is the entire IV scale and *effect* is measured as a percentage figure, e.g., $b_p$=0.1 means a unit effect of 10 percentage points; $b_p$=0.15 indicates 15 points of unit effect, etc. Note that the noun "effect" does not necessarily imply causal effect. Rather, there are predictive effects, expected effects, and causal effects, implying varying degrees of clarity, certainty, or confidence about the possible causal relation between IV and DV. The ability to infer causality is determined by study design, variable selection, and model specification. The choice of measurement scales or statistical indicators does not affect that ability.

**Interpretation 2: Efficiency.** Coefficient $b_p$ represents the *efficiency* with which IV effects DV, where the noun "efficiency" is defined as effect per unit, and the verb "effects" may mean "predicts" or "impacts," indicating varying degrees of causal clarity or confidence as discussed above.

**Interpretation 3: Whole-Scale Effect**. A special feature of percentage scale is that the unit is the entire scale anchored by the conceptual minimum and maximum. Therefore, while $b_p$ represents the *effect per unit*, it also represents the percentage change in DV associated with an increase in IV spanning the entire IV scale, from the conceptual minimum to the conceptual maximum, namely *whole-scale effect* or *entire-scale effect*. Again, the "effect" may imply causal, expected, or predictive effect depending on data and design.

**Interpretation 4: Percent-of-a-point effect (EPOP)**. Coefficient $b_p$ also represents the estimated percent-of-a-point (EPOP) change in DV associated with a one-point increase in IV. Under linear assumption, the "point" can be any point in the DV or the IV scale. Accordingly, $b_p$=0.1 means 10 EPOP, and $b_p$=0.15 indicates 15 EPOP, etc.

In practice researchers have found one expression easier to comprehend or communicate than the others depending on content and context.

The uniform meaning of percentage points allows percentage coefficient ($b_p$) to serve as a generic indicator for comprehension. As a generic indicator, $b_p$ is to aid general, initial, and preliminary comprehension. Coefficient $b_p$ is not meant to be a definitive indicator that is perfect, the best, or the most suitable for understanding every aspect of an effect in every study. We will discuss the limitations or imperfections of $b_p$ as a generic tool for comprehension.

### II.4. Percentage coefficients ($b_p$) are comparable, enabling comparison.

Percentage scales not only encourage interpretation, but also invite comparison. The comparison would assume that *one whole is equitable to another whole* (whole-equals-whole, or WEW, for short). The assumption underlies the common practice of comparing percentage figures. Suppose a researcher observes that Country A has a higher unemployment rate (6%) than Country B (5%), even though the population of Country B is five times larger than that of Country A. The researcher would assume WEW, knowingly or not.

WEW is a philosophical assumption – we hold that whole-equals-whole is a reasonable starting point for analysis. WEW is also a psychological assumption – we assume that people tend to accept WEW as a reasonable starting point. The success of scale normalization, especially min-max normalization of last two decades, buttresses confidence in the WEW assumption. To "normalize scales" means to equalize the scales.

It appears to be common sense among data mining and machine learning researchers that scales need to be normalized, usually at preprocessing stage, otherwise some features will overwhelm others, throwing all subsequent teaching-and-learning off track. (Han et al., 2012; Jain & Bhandare, 2011; Patro & Sahu, 2015).

The WEW assumption enables percentage coefficient ($b_p$) to serve as a generic indicator for comparison. As a generic indicator, $b_p$ is to aid general, initial, and preliminary comparisons. It is not meant to be a definitive indicator that is perfect, the best, or the most suitable for every comparison of every study. We will discuss the limitations or imperfections of $b_p$ as a generic tool for comparison.

## III. A Real Data Example

### III.1. Data

This study employs data from the Health Information National Trends Survey (HINTS 5, Cycle 4), collected between February and June of 2020. The National Cancer Institute (NCI) of the United States initiated HINTS to gather data that is representative of American adults' interactions with health-related information, their health behaviors, and subsequent outcomes. Data collection was conducted via a self-administered postal questionnaire, using a sampling frame from the Marketing System Group's address-based system in the US (see http://hints.cancer.gov/ for further details). This approach employs a robust stratification mechanism, ensuring a sample that accurately reflects the demographic composition of the U.S. population. Of the initial 10,531 individuals contacted, 3,865 completed the survey, resulting in a response rate of 36.7%. The HINTS data encompasses a comprehensive range of demographic, geographic, and psychosocial variables, enabling a detailed examination of the dissemination of health information and its influence on individual health behaviors and outcomes. Such data offers a unique vantage point for evaluating how the public obtains health information, their perception of associated risks, and the use of this information in making informed health decisions.

### III.2. Data Analysis

To facilitate the estimation and comparison of regression coefficients along with their statistical properties, we developed a Python-based tool that employs the percentile bootstrap method. This tool calculates regression coefficients (denoted as $b_p$, $\beta$, and $b_w$) and assesses their standard errors and confidence intervals. The software is available on GitHub at https://github.com/dianshili/Percentage-Coefficients-bp. This algorithm follows the methodology outlined in Hayes (2022).

One mission of developing more precise concepts and more accurate indicators of effect sizes is to reduce the overreliance and mis-reliance on confidence indicators such as *p* values and the so-called statistical significance tests. One pleasant surprise, however, is that more informative effect size measures create legitimate needs for statistical tests that complement and strengthen the effect size assessment and comparison.

In this study, the comparison between percentage coefficients ($b_p$) raise the question whether a difference between two $b_p$ coefficients passes the statistical threshold of $p<\alpha$, where α takes the conventional values of .05, .01, and .001. We applied the percentile bootstrap method to compute the differences between the coefficients and the probability levels of the differences. The computational details are provided in Supplementary Materials (Supplement II) attached to the end of this manuscript. The experimental results are presented in Tables 3, 4, and 5.

### III.3. Findings

The percentage coefficients ($b_p$) in Table 3 and Table 4 indicate the main findings. Raw coefficients ($b_w$) and standardized beta ($\beta$) are provided for comparison. Confidence intervals are also provided to follow the convention. Some of these may be omitted in future studies if $b_p$ is accepted as providing sufficient information.

### III.*3.1. Presentation Order*

The order of presentation follows the following rules.

1) Numerical and binary variables are presented in one batch before nominal variables. (AGE~GEN batch)

2) Any 01-coded (dummy) variable representing missing cases is placed immediately after the corresponding numerical variable (INC & INCmis pair).

3) 01-coded variables representing the same nominal variables are presented as one batch after the numerical-binary batch. (RACwht~RACasn batch).

4) Of the groups represented by a nominal variable, the group showing the lowest or highest DV average serves as the reference group ("Others" group, who showed the highest PSD of the five ethnic groups).

5) Within each batch, the IVs are reverse ordered from high to low by $|b_p|$. For examples, Age (|-.269|) and $RAC_{wht}$ (|-.073|) are on top of their respective batches because each has the largest absolute value, $|b_p|$, in its batch.

### III.*3.2. Main Findings*

**Comprehension.** Percentage coefficient ($b_p$) may be interpreted as percentage change in DV associated with a whole scale increase in IV (Interpretation 1, II.3 above). For example, $b_p$=-.269 (Eq. 9, Table 3) indicates a drop of almost 27-percentage-points in psychological distress (PSD) in association with a 0~100% increase in Age (c.f. E2 & F2, Table 2). Also, $b_p$=-.164 (Eq. 9, Table 3) indicates a 16-points drop in PSD in association with a whole scale increase in education from the lowest (less than eight years of schooling) to the highest (postgraduate) (c.f. E2 & F2, Table 2, and Appendix, 4, EDU).

To percentize the DV scale (PSD) is also to interpret the effect sizes of the 0-1 (dummy) coded IVs, including the binary (INCmis and Gender) and nominal variables (Race). Females, for example, on average reported higher distress than males by 3.4 percentage points ($b_p$=.034, $p$<.001, Line 5 of Eq. 9). The larger-than-threshold $p$-value ($p \geq 0.05$) of INCmis indicates insufficient evidence as to whether those who did not provide valid information about their income had a different PSD level than those who did. Accordingly, we should not further interpret the fact that the former group reported lower PSD ($b_p$=-.022) than the latter group.

Nominal IVs are constructed to investigate the differences between groups. Of $G$ groups, there are $(G^2\text{-}G)/2$ pairwise differences, of which only G-1 01-coded IVs can be represented in a regression equation. This study has 10 pairwise differences, while only four 0-1 coded IVs appear in Table 3. Of all pairwise differences, the largest one needs to be identified and interpreted. The way we chose the reference group (the "Others" group, c.f. IV.3.1, #4) makes this figure readily available as the $b_p$ coefficient with the largest absolute value of the G-1 (four) coefficients in the regression output ($b_p$=-.073, Eq. 9, Table 4).

This coefficient suggests that the White Americans were the least distressed group, who reported 7.3-points lower PSD than the most distressed group, the "Others" group ($b_p$=-.073, $p$<0.01). The Whites were followed by Blacks ($b_p$=-.071, $p$<0.001), Hispanics ($b_p$=-.047, $p$<0.01), and Asian Americans ($b_p$=-.031, $p$<0.05).

Note that the mean percentage score of PSD is 0.17, or 17 points, for all respondents (Table 2, I0), of which 7.3 points would be 42.9% (7.3/17=42.94%). That's a large proportion.

Another figure worth having is the average inter-group difference. This figure, which for this study is the average of the 10 pairwise differences, is not directly available from Eq. 9, but can be easily calculated based on the $b_p$ coefficients available. The result is $\overline{b_p}$=.0372. That would be nearly 22% (.0372/.17=.2188) of the average PSD.

**Comparison.** Coefficients $b_p$ can be compared with each other assuming whole equals whole (WEW, II.4 above), which means the following.

*1. Up-or-Down Comparison*. Of any two $b_p$ coefficients, the larger absolute value indicates larger *effect-per-unit*, aka *efficiency*. For example, Age ($b_p$ = - .269, $p$<.001, Eq. 9, Table 3) is more efficient in reducing Americans' psychological distress than Income ($b_p$ = - .164, $p$<.001). The difference between the two efficiencies (|-.269| - |-.164|=.105) passes the statistical threshold ($p$<.001; Cell A2, Table 4), therefore statistically acknowledged.

*2. Scalar vs Directional Comparisons*. In the physical world, direction and scalar quantity work through their multiplicative product, Euclidean vector, to impact human life (Hoffman, 2013) (Solomentsev, 1994). In parallel, social and psychological valence (direction) and volume (quantity) work through their multiplicative product, *social vector*, to impact human society, to borrow the term from physics and computer science (Chu et al., 2013) (Ivanov, 2001/1994).

The scalar vs direction distinction of the social vector approach guided the construction of Table 4 and Table 5. Table 4 demonstrates scalar comparisons. It displays *scalar differences between efficiencies* ($d_s$), defined as the differences between absolute values of $b_p$ coefficients as shown in Eq. 5, where *i* and *j* identify every IV in rotation.

| $d_s = \|b_p(i)\| - \|b_p(j)\|$ | Eq. 5 |
|---|---|

Table 5 demonstrates directional comparisons. It displays *directional difference between efficiencies* ($d_d$), or simply difference defined as the differences between $b_p$ coefficients based on Eq. 6, where *i* and *j* again identify every IV in rotation.

| $d_d = b_p(i) - b_p(j)$ | Eq. 6 |
|---|---|

Directional comparison differs from scalar comparison only when the regression coefficients bear opposing signs. In this example, all coefficients bear negative signs except that of the binary variable Gender (Table 3). The application of directional comparison would be better demonstrated with a model with opposing coefficients.

*3. Differential vs Proportional Comparisons*. *Differential comparison* calculates the difference between the absolute values of two $b_p$. For example, Age is more efficient than Income by 10.5 percentage points (|-.269| - |-.164|=.105, Eq. 9, Table 3), and is more efficient than Education by 23.4 points (|-.269| - |-.035|=.234).

In contrast, *proportional comparison* calculates the difference between two $b_p$ coefficients as a proportion of the larger or a multiple of the smaller. For example, Age was 61.6% more efficient than Income (0.105/0.164= 0.616) and was almost eight times as efficient as Education (26.9/.035=7.686).

*4. Complement vs Competition Comparisons.* Comparison can be made between two numerical IVs in the same direction to analyze the complementary relation (Jiang et al., 2021). The comparisons of Age, Income, and Education illustrated above are examples.

5. *Competition Comparison.* Comparison can be made between two numerical IVs in opposing directions to analyze the competitive relation (Zhao et al., 2010). An example for competition comparison is not available in Table 3 because all three numerical IVs bear negative signs.

6. *Numerical-Numerical (Num-Num) Comparison*. Comparison can be made between two numerical IVs. The Age-Income and Age-Education are two examples.

7. *Binary-Binary (Bin-Bin) Comparison*. A comparison can be made between two binary IVs. Comparing Gender and $INC_{mis}$ (missing values of income), one may see that the bp coefficients show a difference of 1.2 percentage points ($b_p$=.034 vs $b_p$=-.022, Eq. 9, Table 3). The difference is small considering the natures of the two IVs, even though the difference is statistically acknowledged ($p < .01$, E3 of Table 4).

8. *Numerical-Binary (Num-Bin) Comparison*. Comparison can be made between a numerical IV and a binary IV. While education is negatively correlated with PSD ($b_p$ = - .035, $p$<.05), the difference between the least and the most educated is slightly larger than that between the two genders ($b_p$=.034, $p$<.001). The difference is small (|-.035| - |.034|=.002, with rounding error) despite passing the statistical threshold ($p$<.001, D5 of Table 4).

9. *Numerical-Nominal (Num-Nom) Comparison*. As said, of a nominal variable represented by a set of 0-1 (dummy) variables, two figures need noticing. One is $b_p$ coefficient representing the largest intergroup difference. In this study, it is $b_p$=-.073, representing the White-Others difference. The second figure is the $|b_p|$ averaged across all possible pairs. In this study, it is $\overline{|b_p|}$=.0372, representing the average across 10 pairwise differences (($5^2$-5)/2=10).

The figures serve as good reference points to be compared with the $b_p$ of a numerical IV. Comparing $b_p$ =-.164 (INC) with $b_p$=-.073 (White-Others), we see that the difference in PSD between the lowest and the highest income earners was more than twice the difference between the Whites and the “Others” group (|-.164|/|-.073|=2.247, Eq. 9, Table 3. $p$<.01, B6, Table 4). Given that the White-Others difference is the largest of all differences between ethnic groups, the comparison also implies that income more than matches the efficiency of ethnicity when predicting psychological distress.

By contrast, while Education and Ethnicity both affect psychological distress (Line 4 and Lines 7~10, Table 3), there is no evidence that the two sets of effects differ from each other in magnitudes. The four $d_s$ differentials all fail to pass the statistical threshold ($p$≥0.05, D6~D9, Table 4). Furthermore, the absolute values of the differentials appear small, especially when compared to that of Age (A6~A9) or Income (B6~B9).

While the Gender-Ethnicity differentials (|$d_s$| of E6-E9) appear as small as the Education-Ethnicity differentials (|$d_s$| of D6-D9), each of the former passed the statistical threshold ($p$<.001) while each of the latter failed ($p$≥.05). It is strong evidence for a small edge of the ethnicity effect over the gender effect, and weak evidence for any edge of the ethnicity effect over the education effect. Methodology task forces and experts have warned for decades that smaller $p$-values must not be confused with larger effects or greater significance. The case analyzed here constitutes yet another empirical example in support of the warnings.

**III.4. Roles of Other Indicators**

### ***Raw Coefficient ($b_w$) and Standardized Beta ($\beta$)***

Raw *b* ($b_w$) and standardized beta ($\beta$) were also provided in Table 3 as they would be in typical studies of multiple regression. As usual, the raw *b* coefficients ($b_w$) are difficult to interpret because the 1-4 raw DV scale does not offer clearly and uniformly meaningful units. The same is true with Income and Education, the two IVs on close-end scales (Appendix and Table 2). The $b_w$ coefficients also cannot be meaningfully compared with each other. That's because, other than the 0-or-1 binary variables, the units of the IV scales are not equitable with each other.

As usual, the $\beta$ coefficients cannot be interpreted or compared substantively because standardized deviations, as scale units, do not offer substantive meanings and cannot be said to be substantively equitable to each other.

### ***P-values and Significance Tests***

This study adopts the practice of employing *p*-values and NHST "significance tests" as pretests, passing which would allow researchers to interpret and compare the effect sizes with above-threshold confidence. (Han et al., 2023; Jiang et al., 2021; Liu, Chang, et al., 2023; Zhao, Feng, et al., 2022; Zhao & Zhang, 2014). The study also adopts the view that "significance," including "statistical significance" and its variations and derivations, is a misnomer, therefore needs to be avoided if feasible. (Amrhein et al., 2019; Correia et al., 2019; Wasserstein & Lazar, 2016; Wasserstein et al., 2019). "Significance-less phrases," such as "statistical acknowledgement" for $p < 0.05$ and "statistically inconclusive" for $p \geq 0.05$, are used instead. (Han et al., 2023; Jiang et al., 2021; Zhao, Feng, et al., 2022).

### ***Confidence Intervals***

Confidence intervals are provided following the convention. Now that $b_p$ provides comprehendible and comparable information about the effect sizes, and *p*-asterisks indicate the confidence levels about the effects' directions, the burden on the confidence levels to provide additional information is reduced.

## IV. Conclusion: $b_p$ as a Generic Indictor of Efficiency as a Facet of Effect

Percentage coefficients are conceived and constructed as a generic indicator of efficiency, i.e., unit effect. The definition implies the following:

1) Coefficient $b_p$ is not meant to be the definitive or encompassing indicator of effect for every and all purposes. Rather, it is generic indicator of efficiency, aka effect per unit, where "effect" may imply predictive effect, expected effect, or causal effect depending on the nature of the data.

2) As a *generic* indicator, $b_p$ serves as a quick and convenient indicator of efficiency useful in many situations. More importantly, it may serve as stimulation and reference for more targeted measures designed for specified purposes of specified studies. It is like an over-the-counter generic drug conveniently available for quick remedy in many situations. Prescription or brand-named medicines may provide more pinpointed, prompt, or potent remedies.

1) Therefore, for any IV-DV relationship, $b_p$ should not be the conclusive end of measuring the effect, but rather a stimulating beginning to converse and collaborate for more accurate and appropriate measures.

Table 2. Variable Percentization ($n$ =3,865).

| | | raw scores ($s_r$) (observed) | | | | conceptual range | | 01 percentage scores ($s_p$) (observed) | | | |
|---|---|---|---|---|---|---|---|---|---|---|---|
| | | Min | Max | Mean | SD | Min ($c_n$) | Max ($c_x$) | Min | Max | Mean | SD |
| 0. | PSD | 1 | 4 | 1.50 | 0.73 | 1 | 4 | 0.00 | 1.00 | 0.17 | 0.24 |
| 1. | AGE | 18 | 104 | 57.01 | 16.99 | 0 | 100 | 0.18 | 1.04 | 0.57 | 0.17 |
| 2. | INC | 1 | 9 | 5.59 | 2.13 | 1 | 9 | 0.00 | 1.00 | 0.57 | 0.27 |
| 3. | $INC_{mis}$ | 0 | 1 | 0.11 | 0.31 | 0 | 1 | 0.00 | 1.00 | 0.11 | 0.31 |
| 4. | EDU | 1 | 7 | 4.93 | 1.62 | 1 | 7 | 0.00 | 1.00 | 0.66 | 0.27 |
| 5. | GEN | 0 | 1 | 0.59 | 0.49 | 0 | 1 | 0.00 | 1.00 | 0.59 | 0.49 |
| 6. | $RAC_{otr}$ | 0 | 1 | 0.13 | 0.33 | 0 | 1 | 0.00 | 1.00 | 0.13 | 0.33 |
| 7. | $RAC_{wht}$ | 0 | 1 | 0.55 | 0.50 | 0 | 1 | 0.00 | 1.00 | 0.55 | 0.50 |
| 8. | $RAC_{blk}$ | 0 | 1 | 0.12 | 0.33 | 0 | 1 | 0.00 | 1.00 | 0.12 | 0.33 |
| 9. | $RAC_{hsp}$ | 0 | 1 | 0.15 | 0.36 | 0 | 1 | 0.00 | 1.00 | 0.15 | 0.36 |
| 10. | $RAC_{asn}$ | 0 | 1 | 0.04 | 0.20 | 0 | 1 | 0.00 | 1.00 | 0.04 | 0.20 |

Table 3: Effects of Demographic Variables Predicting Psychological Distress

| | Eq. 7 | Eq. 8 | Eq. 9 |
|---|---|---|---|
| Right: Dependent Variable (DV) | PSD Psychological Distress | PSD Psychological Distress | PSD Psychological Distress |
| Right: effect size indicators | raw *b* $b_w$ | standardized beta $\beta$ | percentage coefficient $b_p$ |
| 0. Intercept | 2.454 [2.304, 2.604] *** | .003 [-0.028, 0.035] | .458 [0.411, 0.505] *** |
| 1. AGE | -.008 [-0.009, -0.007] *** | -.189 [-0.222, -0.156] *** | -.269 [-0.316, -0.222] *** |
| 2. INC | -.062 [-0.074, -0.049] *** | -.181 [-0.216. -0.145] *** | -.164 [-0.196, -0.132] *** |
| 3. $INC_{mis}$ (miss=1, no miss=0) | -.065 [-0.155, 0.025] | -.028 [-0.067, 0.011] | -.022 [-0.052, 0.008] |
| 4. EDU | -.018 [-0.034, -0.001] * | -.040 [-0.076, -0.003] * | -.035 [-0.068, -0.003] * |
| 5. GEN (female=1, male=0) | .101 [0.054, 0.147] *** | .068 [0.037, 0.100] *** | .034 [0.018, 0.049] *** |
| 6. $RAC_{otr}$ (Others as reference) | | | |
| 7. $RAC_{wht}$ (White vs Others) | -.218 [-0.350, -0.085] ** | -.060 [-0.097, -0.023] ** | -.073 [-0.117, -0.028] ** |
| 8. $RAC_{blk}$ (Black vs Others) | -.214 [-0.312, -0.115] *** | -.097 [-0.142, -0.052] *** | -.071 [-0.104, -0.038] *** |
| 9. $RAC_{hsp}$ (Hispanic vs others) | -.141 [-0.237, -0.045] ** | -.070 [-0.118, -0.023] ** | -.047 [-0.079, -0.015] ** |
| 10. $RAC_{asn}$ (Asians vs others) | -.094 [-0.176, -0.012] * | -.064 [-0.120, -0.008] * | -.031 [-0.059, -0.004] * |
| Total $r^2$ | .075 *** | .075 *** | .075 *** |

* p<.05; ** p<.01; *** p<.001.

Main cells are regression coefficient, $b_w$, $\beta$, or $b_p$, and (confidence intervals).

*Raw coefficient* ($b_w$) is regression coefficients on raw scales that has not been sufficiently processed for comprehension or comparison.

*Standardized coefficient* ($\beta$) is regression coefficient on standardized scales.

*Percentage coefficient* ($b_p$) is regression coefficient on 01 percentage scales. It indicates the percentage change in DV associated with a whole scale increase in IV. Under linear assumption, it also indicates a percent-of-one-point (POP) change in DV associated with a one-point increase in IV.

Raw scales, i.e., the unprocessed or insufficiently processed scales that data analysts encounter initially, often feature units that do not offer uniform meaning. Likert scales and other ordinal scales are typical examples. When the unit of a scale is not uniformly meaningful, *raw coefficient* ($b_w$) based on the scale is uninterpretable, therefore incomprehensible, and incomparable.

**List of abbreviations:**

PSD: Psychological Distress
GEN: Gender-Female
AGE: Age
EDU: Education
INC: Household Income
$INC_{mis}$: Dummy to represent missing cases of INC[①]

---

① Further readings for dummy variable adjustment:
1. https://statisticalhorizons.com/is-dummy-variable-adjustment-ever-good-for-missing-data/
2. Cohen, J., Cohen, P., West, S. G., & Aiken, L. S. (2003). Applied multiple regression/correlation analysis for the behavioral sciences, 3rd Ed. Lawrence Erlbaum

RAC: Race

$RAC_{wht}$: Non-Hispanic White

$RAC_{blk}$: Black or African American

$RAC_{hsp}$: Hispanic

$RAC_{asn}$: Non-Hispanic Asian

$RAC_{otr}$: Non-Hispanic Other

Table 4: Scalar Comparison of Efficiencies, $d_s=|b_p(i)| - |b_p(j)|$

| | | A | B | C | D | E | F | G | H | I |
|---|---|---|---|---|---|---|---|---|---|---|
| | *j* \ *i* | **AGE** | **INC** | **$INC_{mis}$** | **Edu** | **GEN** | **$RAC_{wht}$** | **$RAC_{blk}$** | **$RAC_{hsp}$** | **$RAC_{asn}$** |
| 1 | **AGE** | -- | −0.105*** | −0.247*** | −0.233*** | −0.235*** | −0.196*** | −0.197*** | −0.222*** | −0.237*** |
| 2 | **INC** | 0.105*** | -- | −0.142*** | −0.129*** | −0.131*** | −0.091** | −0.093*** | −0.117*** | −0.133*** |
| 3 | **$INC_{mis}$** | 0.247*** | 0.142*** | -- | 0.014 | 0.012** | 0.051 | 0.050* | 0.025 | 0.010 |
| 4 | **Edu** | 0.233*** | 0.129*** | −0.014 | -- | −0.002*** | 0.037 | 0.036 | 0.012 | −0.004 |
| 5 | **GEN** | 0.235*** | 0.131*** | −0.012** | 0.002*** | -- | 0.039*** | 0.038*** | 0.014*** | −0.002*** |
| 6 | **$RAC_{wht}$** | 0.196*** | 0.091** | −0.051 | −0.037 | −0.039*** | -- | −0.001 | −0.026 | −0.041* |
| 7 | **$RAC_{blk}$** | 0.197*** | 0.093*** | −0.050* | −0.036 | −0.038*** | 0.001 | -- | −0.024 | −0.040** |
| 8 | **$RAC_{hsp}$** | 0.222*** | 0.117*** | −0.025 | −0.012 | −0.014*** | 0.026 | 0.024 | -- | −0.016 |
| 9 | **$RAC_{asn}$** | 0.237*** | 0.133*** | −0.010 | 0.004 | 0.002*** | 0.041* | 0.040** | 0.016 | -- |

Main cells are *scalar differences between efficiencies* ($d_s$), $d_s=|b_p(i)| - |b_p(j)|$. The $b_p$ coefficients are shown in Table 3.
* $p<.05$, ** $p<.01$, *** $p<.001$.

Table 5: Directional Comparison of Efficiencies, $d_d = b_p(i) - b_p(j)$)

| | | A | B | C | D | E | F | G | H | I |
|---|---|---|---|---|---|---|---|---|---|---|
| | *j* \ *i* | AGE | INC | $INC_{mis}$ | Edu | GEN | $RAC_{wht}$ | $RAC_{blk}$ | $RAC_{hsp}$ | $RAC_{asn}$ |
| 1 | AGE | -- | 0.105*** | 0.247*** | 0.233*** | 0.302*** | 0.196*** | 0.197*** | 0.222*** | 0.237*** |
| 2 | INC | −0.105*** | -- | 0.142*** | 0.129*** | 0.198*** | 0.091** | 0.093*** | 0.117*** | 0.133*** |
| 3 | $INC_{mis}$ | −0.247*** | −0.142*** | -- | −0.014 | 0.055** | −0.051 | −0.050* | −0.025 | −0.010 |
| 4 | Edu | −0.233*** | −0.129*** | 0.014 | -- | 0.069*** | −0.037 | −0.036 | −0.012 | 0.004 |
| 5 | GEN | −0.302*** | −0.198*** | −0.055** | −0.069*** | -- | −0.106*** | −0.105*** | −0.081*** | −0.065*** |
| 6 | $RAC_{wht}$ | −0.196*** | −0.091** | 0.051 | 0.037 | 0.106*** | -- | 0.001 | 0.026 | 0.041* |
| 7 | $RAC_{blk}$ | −0.197*** | −0.093*** | 0.050* | 0.036 | 0.105*** | −0.001 | -- | 0.024 | 0.040** |
| 8 | $RAC_{hsp}$ | −0.222*** | −0.117*** | 0.025 | 0.012 | 0.081*** | −0.026 | −0.024 | -- | 0.016 |
| 9 | $RAC_{asn}$ | −0.237*** | −0.133*** | 0.010 | −0.004 | 0.065*** | −0.041* | −0.040** | −0.016 | -- |

Main cells are (*directional) difference between efficiencies* ($d_d$), $d_d = b_p(i) - b_p(j)$. The $b_p$ coefficients are shown in Table 3.
* $p$<.05, ** $p$<.01, *** $p$<.001.

# References

Achen, C. H. (1982). *Interpreting and using regression* (Vol. 29). Sage.

AERA_Task_Force. (2006). Standards for reporting on empirical social science research in AERA publications, by AERA Task Force on Reporting of Research Methods in AERA Publications. *Educational researcher*, *35*, 33-40. https://doi.org/10.3102/0013189X035006033

Afifi, A. A., & Clarke, V. (1990). *Computer Aided Multivariate Analysis* (2 ed.). Van Nostrand Reinhold.

Amrhein, V., Greenland, S., McShane, B., & others. (2019). Retire statistical significance. *Nature*, *567*, 305-307. https://doi.org/10.1038/d41586-019-00857-9

Ao, S. H., Zhang, L., Liu, P. L., & ZHAO, X. (2023). Social Media and Partnership Jointly Alleviate Caregivers' Psychological Distress: Exploring the Effects of Online and Offline Connectedness. *BMC Psychology*.

Baker, M. (2016). 1,500 scientists lift the lid on reproducibility. *Nature*, *533*(7604), 452-454. https://doi.org/10.1038/533452a

Benjamini, Y., Veaux, R. D., Efron, B., Evans, S., Glickman, M., Graubard, B. I., He, X., Meng, X.-L., Reid, N., Stigler, S. M., Vardeman, S. B., Wikle, C. K., Wright, T., Young, L. J., & Kafadar, K. (2021). ASA President's Task Force Statement on Statistical Significance and Replicability. *Harvard Data Science Review*. https://doi.org/10.1162/99608f92.f0ad0287

Berkson, J. (1938). Some difficulties of interpretation encountered in the application of the chi-square test. *Journal of the American Statistical Association*, *33*, 526- 536. https://doi.org/10.1080/01621459.1938.10502329

Bring, J. (1994). How to standardize regression coefficients. *The American Statistician*, *48*(3), 209-213.

Celsius temperature scale. (2023). In *Encyclopædia Britannica Online*.

Chin, J. M., Pickett, J. T., Vazire, S., & Holcombe, A. O. (2023). Questionable research practices and open science in quantitative criminology. *Journal of Quantitative Criminology*, *39*(1), 21-51.

Chu, T. H. S., Hui, F. C. P., & Chan, H. C. B. (2013). Smart Shopping System Using Social Vectors and RFID. *International MultiConference of Engineers and Computer Scientists 2013*, *I*.

Cohen, J. (1990). Things I have learned (so far). *American psychologist*, *45*, 1304-1312. https://doi.org/10.1037/10109-028

Cohen, J. (1994). The earth is round (p<. 05). *American Psychologist*, *49*, 997-1003. https://doi.org/10.1037/0003-066X.49.12.997

Collaboration, O. S. (2015). Estimating the reproducibility of psychological science. *Science*, *349*(6251), aac4716.

Correia, L. C. L., Bagano, G. O., & Melo, M. H. V. d. (2019). Should we retire statistical significance? *Brazilian Journal Of Pain*, *2*(3). https://doi.org/10.5935/2595-0118.20190037

Darlington, R. B. (1968). Multiple regression in psychological research and practice. *Psychological bulletin*, *69*(3), 161.

Darlington, R. B. (1990). *Regression and linear models* McGraw-Hill.

Fahrenheit Gabriel Daniel. (1911). In H. Chisholm (Ed.), *Encyclopædia Britannica.* (11 ed., Vol. 10, pp. 126). Cambridge: Cambridge University Press.

Fahrenheit temperature scale. (2023). In *Encyclopædia Britannica Online*.

Greenland, S., Maclure, M., Schlesselman, J. J., Poole, C., & Morgenstern, H. (1991). Standardized regression coefficients: a further critique and review of some alternatives. *Epidemiology*, *2*(5), 387-392.

Greenland, S., Schlesselman, J. J., & Criqui, M. H. (1986). The fallacy of employing standardized regression coefficients and correlations as measures of effect. *American Journal of Epidemiology*, *123*(2), 203-208. https://doi.org/10.1093/oxfordjournals.aje.a114229

Halsey, L. G., Curran-Everett, D., Vowler, S. L., & Drummond, G. B. (2015). The fickle P value generates irreproducible results. *Nature Methods*, *12*(3), 179-185.

Han, J., Kamber, M., & Pei, J. (2012). *Data Mining: Concepts and Techniques* (3 ed.). Morgan Kaufmann - Elsevier.

Han, J. K., & Kamber, M. (2001). *Data Mining: Concepts and Techniques*. Morgan Kaufmann Publishers.

Han, T., Zhang, L., Zhao, X., & Deng, K. (2023). Total-effect Test May Erroneously Reject So-called "Full" or "Complete" Mediation. *arXiv preprint arXiv:2309.08910*. https://arxiv.org/pdf/2309.08910.pdf

Hanushek, E. A., & Jackson, J. E. (1978). *Statistical methods for social scientists*. Academic Press.

Hanushek, E. A., & Jackson, J. E. (2013). *Statistical methods for social scientists*. Academic Press.

Hayes, A. F. (2022). *Introduction to mediation, moderation, and conditional process analysis: A regression-based approach*. The Guilford Press.

He, X., Madigan, D., Yu, B., & Wellner, J. (2019). Statistics at a Crossroads: Who is for the Challenge? - Report 2019 for National Science Foundation. National Science Foundation Workshop.,

Head, M. L., Holman, L., Lanfear, R., Kahn, A. T., & Jennions, M. D. (2015). The extent and consequences of p-hacking in science. *PLoS biology*, *13*(3), e1002106.

Hoffman, K. (2013). *Analysis in Euclidean space*. Courier Corporation.

Huang, H. (2023). Statistics reform: practitioner's perspective. *ResearchGate Preprint*, 1-28. https://doi.org/10.13140/RG.2.2.31799.50084

Ivanov, A. B. (2001/1994). Vector. In *Encyclopedia of Mathematics*.

Jain, A. K., Nandakumar, K., & Ross, A. a. (2005). Score Normalization in Multimodal Biometric Systems. *Pattern recognition*, *38*, 2270-2285. https://doi.org/10.1016/j.patcog.2005.01.012

Jain, Y. K., & Bhandare, S. K. (2011). Min max normalization based data perturbation method for privacy protection. *International Journal of Computer & Communication Technology*, *2*(8), 45-50.

Jasny, B., Wigginton, N., McNutt, M., Bubela, T., Buck, S., Cook-Deegan, R., Gardner, T., Hanson, B., Hustad, C., & Kiermer, V. (2017). Fostering reproducibility in industry-academia research. *Science*, *357*(6353), 759-761.

Jiang, Y., Zhao, X., Zhu, L., Liu, J. S., & Deng, K. (2021). Total-effect test is superfluous for establishing complementary mediation. *Statistica Sinica*, 1961-1983. https://doi.org/10.5705/ss.202019.0150

Kelley, K., & Preacher, K. J. (2012). On effect size. *Psychological methods*, *17*(2), 137-152. https://doi.org/10.1037/a0028086

Kim, J.-O., & Ferree Jr, G. D. (1981). Standardization in causal analysis. *Sociological Methods & Research*, *10*(2), 187-210.

King, G. (1986). How not to lie with statistics: Avoiding common mistakes in quantitative political science. *American Journal of Political Science*, 666-687.

Kirk, R. E. (1996). Practical significance: A concept whose time has come. *Educational and psychological measurement*, *56*, 746-759. https://doi.org/10.1177/0013164496056005002

Liu, J., & Li, D. (2023). Is Machine Learning Unsafe and Irresponsible in Social Sciences? Paradoxes and Reconsidering from Recidivism Prediction Tasks. *arXiv preprint arXiv:2311.06537*.

Liu, P. L., Chang, A., Liu, M. T., Ye, J. F., Jiao, W., Ao, H. S., Hu, W., Xu, K., & Zhao, X. (2023). Effect of Information Encounter on Concerns over Healthy Eating – Mediated through Body Comparison and Moderated by Body Mass Index or Body Satisfaction. *BMC Public Health*, *23*(254), 1-13. https://doi.org/10.1186/s12889-023-15069-0

Liu, P. L., Ye, J. F., Ao, H. S., Sun, S., Li, Q., Zheng, Y., Feng, G. C., Wang, H., & Zhao, X. (2023). Effects of Electronic Personal Health Information (ePHI) Technology on American Women's Cancer Screening Behaviors Mediated through Cancer Worry: Differences and Similarities between 2017 and 2020. *Digital Health*, *9*, 1-12. https://doi.org/10.1177/20552076231185271

Loftus, G. R. (1991). On the Tyranny of Hypothesis Testing in the Social Sciences. *Contemporary Psychology: A Journal of Reviews*, *36*(2), 102-105. https://doi.org/10.1037/029395

Loftus, G. R. (1996). Psychology will be a much better science when we change the way we analyze data. *Current directions in psychological science*, *5*(6), 161-171.

Lykken, D. T. (1968). Statistical significance in psychological research. *Psychological bulletin*, *70*(3p1), 151.

Makel, M. C., Plucker, J. A., & Hegarty, B. (2012). Replications in psychology research: How often do they really occur? *Perspectives on Psychological Science*, *7*(6), 537-542.

Masicampo, E., & Lalande, D. R. (2012). A peculiar prevalence of p values just below. 05. *Quarterly journal of experimental psychology*, *65*(11), 2271-2279.

Matthews, R. (2021). The p-value statement, five years on. *Significance*, *18*, 16-19. https://doi.org/10.1111/1740-9713.01505

Matthews, R., Wasserstein, R., & Spiegelhalter, D. (2017). The ASA's p-value Statement, One Year on. *Significance*, *14*(2), 38-41. https://doi.org/10.1111/j.1740-9713.2017.01021.x

Meehl, P. E. (1954). *Clinical versus statistical prediction: A theoretical analysis and a review of the evidence*. University of Minnesota Press. https://doi.org/10.1037/11281-000

Meehl, P. E. (1992). Theoretical risks and tabular asterisks: Sir Karl, Sir Ronald, and the slow progress of soft psychology.

Munafò, M. R., Nosek, B. A., Bishop, D. V., Button, K. S., Chambers, C. D., Percie du Sert, N., Simonsohn, U., Wagenmakers, E.-J., Ware, J. J., & Ioannidis, J. (2017). A manifesto for reproducible science. *Nature human behaviour*, *1*(1), 1-9.

Nuzzo, R. (2014). Statistical Errors: P values, the 'gold standard' of statistical validity, are not as reliable as many scientists assume. *Nature*, *506*, 150-152. https://doi.org/10.1038/506150a

Patro, S., & Sahu, K. K. (2015). Normalization: A preprocessing stage. *arXiv preprint arXiv:1503.06462*.

Pedhazur, E. J., & Kerlinger, F. N. (1982). *Multiple regression in behavioral research: Explanation and prediction* (2 ed.). Holt, Rinehart & Winston.

Preacher, K. J., & Kelley, K. (2011). Effect size measures for mediation models: quantitative strategies for communicating indirect effects. *Psychological methods*, *16*(2), 93-115. https://doi.org/10.1037/a0022658

Robinson, D. H., Fouladi, R. T., Williams, N. J., & Bera, S. J. (2002). Some Effects of Including Effect Size and "what if" Information. *The Journal of Experimental Education*, *70*(4), 365-382. https://doi.org/10.1080/00220970209599513

Robinson, D. H., Whittaker, T. A., Williams, N. J., & Beretvas, S. N. (2003). It's Not Effect Sizes So Much as Comments About Their Magnitude That Mislead Readers. *The Journal of Experimental Education*, *72*(1), 51-64. https://doi.org/10.1080/00220970309600879

Shalabi, L. A., Shaaban, Z., & Kasasbeh, B. (2006). Data Mining: A Preprocessing Engine. *Journal of Computer Science*, *2*(9), 735-739.

Simmons, J. P., Nelson, L. D., & Simonsohn, U. (2011). False-positive psychology: Undisclosed flexibility in data collection and analysis allows presenting anything as significant. *Psychological science*, *22*(11), 1359-1366.

Snyder, P., & Lawson, S. (1993). Evaluating results using corrected and uncorrected effect size estimates. *The Journal of Experimental Education*, *61*(4), 334-349.

Solomentsev, E. D. (1994). Euclidean space. In *Encyclopedia of Mathematics*: EMS Press.

Thompson, B. (1996). AERA Editorial Policies Regarding Statistical Significance Testing: Three Suggested Reforms. *Educational researcher*, *25*(2), 26-30.

Thompson, B. (1998). Statistical significance and effect size reporting: Portrait of a possible future. *Research in the Schools*, *5*(2), 33-38.

Thompson, B. (1999). Journal editorial policies regarding statistical significance tests: Heat is to fire as p is to importance. *Educational Psychology Review*, *11*, 157-169. https://doi.org/Doi 10.1023/A:1022028509820
Thompson, B. (2004). The "significance" crisis in psychology and education. *Journal of Socio-Economics*, *33*, 607-613. https://doi.org/10.1016/j.socec.2004.09.034
Thorson, E., Friestad, M., & Zhao, X. (1987). *Attention to program context in a natural viewing environment: Effects on memory and attitudes toward commercials* Association for Consumer Research, Boston,
Television viewing behavior as an indicator of commercial effectiveness, Measuring Advertising Effectiveness 221-238 (Lawrence Erlbaum 1997).
Trafimow, D. (2018). An a priori solution to the replication crisis. *Philosophical Psychology*, *31*, 1188-1214. https://doi.org/10.1080/09515089.2018.1490707
Trafimow, D., & Marks, M. (2015). Editorial. *Basic and Applied Social Psychology*, *37*(1), 1-2. https://doi.org/10.1080/01973533.2015.1012991
Tukey, J. W. (1954). Causation, regression, and path analysis. In O. Kempthorne, T. A. Bancroft, J. W. Gowen, & J. L. Lush (Eds.), *Statistics and mathematics in biology* (pp. 35-66). Iowa State College Press.
Walsh, A. (1990). *Statistics for the social sciences: with computer applications*. Harper & Row.
Wasserstein, R. L., & Lazar, N. A. (2016). The ASA's statement on p -Values: context, process, and purpose. *American Statistician*, *70*, 129-133. https://doi.org/10.1080/00031305.2016.1154108
Wasserstein, R. L., Schirm, A. L., & Lazar, N. A. (2019). Moving to a world beyond “p < 0.05”. *American Statistician*, *73*, 1-19. https://doi.org/10.1080/00031305.2019.1583913
Wilkinson, L., & APA_Task_Force. (1999). Statistical methods in psychology journals: Guidelines and explanations (Report by Task Force on Statistical Inference, APA Board of Scientific Affairs). *American Psychologist*, *54*, 594-604. https://doi.org/10.1037/0003-066X.54.8.594
Wooditch, A., Fisher, R., Wu, X., & Johnson, N. J. (2020). P-value problems? An examination of evidential value in criminology. *Journal of Quantitative Criminology*, *36*, 305-328.
Woolston, C. (2015). Psychology journal bans P values. *Nature*, *519*(7541), 9. https://doi.org/10.1038/519009f
Zhang, L., Ao, S. H., Ye, J. F., & Zhao, X. (2023). How Does Health Communication on Social Media Influence E-cigarette Perception and Use? A Trend Analysis from 2017 to 2020. *Addictive Behaviors*. https://doi.org/10.1016/j.addbeh.2023.107875
Zhang, L., Ao, S. H., & Zhao, X. (2023). Longitudinal relationship between social media and e-cigarette use among adolescents: the roles of internalizing problems and academic performance. *BMC Public Health*, *23*(1). https://doi.org/10.1186/s12889-023-17059-8
Zhao, X. (1989). *Effects of Commercial Position in Television Programming* University of Wisconsin--Madison].
Zhao, X. (1997). Clutter and serial order redefined and retested. *Journal of Advertising Research*, *37*, 57-73.
Zhao, X., Feng, G. C., Ao, S. H., & Liu, P. L. (2022). Interrater reliability estimators tested against true interrater reliabilities. *BMC Medical Research Methodology*, *22*(232), 1-19. https://doi.org/10.1186/s12874-022-01707-5
Zhao, X., Liu, X., Chen, Y. S., Jiao, W. A., Ao, S. H., Shen, F., & Zhao, Z. G. (2023). First-Person Influences on Third-Person Perceptions. *China Media Research*. https://repository.um.edu.mo/handle/10692/126619,

https://www.chinamediaresearch.net/

Zhao, X., Lynch, J. G., & Chen, Q. (2010). Reconsidering Baron and Kenny: Myths and truths about mediation analysis. *Journal of Consumer Research*, *37*, 197-206. https://doi.org/https://doi.org/10.1086/651257

Zhao, X., & Xie, Y. (1992). Western influence on (People's Republic of China) Chinese students in the United States. *Comparative Education Review*, *36*, 509. https://doi.org/10.1086/447148

Zhao, X., Ye, J., Sun, S., Zhen, Y., Zhang, Z., Xiao, Q., Ao, S., Feng, G., & Li, X. (2022). Best Title Lengths of Online Postings for Highest Read and Relay. *Journalism and Communication Review*, *75*, 5-20. https://doi.org/10.14086/j.cnki.xwycbpl.2022.03.001

Zhao, X., & Zhang, X. J. (2014). Emerging methodological issues in quantitative communication research. In J. Hong (Ed.), *New Trends in Communication Studies, II* (pp. 953-978). Tsinghua University Press. https://repository.um.edu.mo/handle/10692/37055

Zhao, X., Zhu, J. H., Li, H., & Bleske, G. L. (1994). Media effects under a monopoly: The case of Beijing in economic reform. *International Journal of Public Opinion Research*, *6*, 95-117. https://doi.org/10.1093/ijpor/6.2.95

Zientek, L. R., Capraro, M. M., & Capraro, R. M. (2016). Reporting Practices in Quantitative Teacher Education Research: One Look at the Evidence Cited in the AERA Panel Report. *Educational researcher*, *37*(4), 208-216. https://doi.org/10.3102/0013189x08319762

# Supplementary Materials

## Supplement I

## Measurement Wording and Variable Construction

**1. PSD,** *psychological distress,* is based on four questions below (Cronbach's α =.88). The four 1~4 items were taken the average to construct PSD ranging 1~4, where 1 indicates that a respondent did not feel at all any of the symptoms, and 4 indicates he or she felt all four symptoms nearly every day (mean=1.50, SD=0.73).

| Q. H8: The past 2 weeks how often have you been bothered by: | Not at all | Several days | More than half | Nearly every day | Non Valid | Total |
|---|---|---|---|---|---|---|
| Right: coding | 1 | 2 | 3 | 4 | Miss. | |
| H8a. Little interest or pleasure in doing things? | 65.5 | 19.6 | 6.8 | 5.6 | 2.5 | 100 |
| H8b. Feeling down, depressed or hopeless? | 70.1 | 18.1 | 5.7 | 3.4 | 2.7 | 100 |
| H8c. Feeling nervous, anxious or on edge? | 60.7 | 24.8 | 6.3 | 5.7 | 2.5 | 100 |
| H8d. Not being able to stop or control worrying? | 67.3 | 18.3 | 6.2 | 5.8 | 2.5 | 100 |

**2.** **RAC**, *race,* is coded as four dummy variables, i.e., $Rac_{wht}$ (white), $Rac_{blk}$ (African American), $Rac_{hsp}$ (Hispanic), $Rac_{asn}$ (Asian), with "others," including other races and no-answers, as the reference group.

Q. P9. What is your race?

| Race | White | Black | Hispanic | Asian | Others | | Total |
|---|---|---|---|---|---|---|---|
| | | | | | Other races | No Answer | |
| Frq. (%) | 55.2 | 12.4 | 15.4 | 4.2 | 3.1 | 9.7 | 100 |

**4.** **EDU**, *education*. Q. P7: Highest grade or school? This variable is coded as the estimated years of schooling within range. (Mean=4.94, SD=1.62).

| response range | < 8 years | 8~11 | 12 yrs / hi sch | post high school | some college | college graduate | post-graduate | non valid | Total |
|---|---|---|---|---|---|---|---|---|---|
| Coding | 1 | 2 | 3 | 4 | 5 | 6 | 7 | mis. | |
| Frq.(%) | 2.1 | 5.0 | 18.2 | 6.8 | 21.1 | 25.3 | 17.7 | 3.7 | 100 |

**5.** **INC**: Household income. Q. P16. This variable is coded as the best estimation within range (Mean=5.59, SD=2.13), as shown below.

| response range | < 10K | 10~ <15K | 15~ <20K | 20~ <35K | 35~ <50K | 50~ <75K | 75~ <100K | 100~ <200K | > 200K | Non valid | To-tal |
|---|---|---|---|---|---|---|---|---|---|---|---|
| Coding | 1 | 2 | 3 | 4 | 5 | 6 | 7 | 8 | 9 | dmy. | |
| Frq. (%) | 5.9 | 5.1 | 5.1 | 11.7 | 11.9 | 15.3 | 10.4 | 17.8 | 6.0 | 10.8 | 100 |

**6. AGE.** Q. P1. What is your age? (Mean=57.0, SD=17.0, Min=18, Max=104). Unit of natural scale (UNS) is *year*.

**7. Gender-Female (GEN)** is coded as 1 for female (57.0%) and 0 for male.

Q. P2: On your original birth certificate, were you listed as male or female?

# Supplement II

## Bootstrap Statistical Tests for Comparing $b_p$ Coefficients

We applied the percentile bootstrap method to compute the differences between efficiencies, as detailed in Equations 5 and 6. The pseudo-code for our approach is presented below:

---

**Input:**

**Dataset D**: Dataset from which samples are drawn.

**n_*bootstrap***: Number of bootstrap samples to generate.

**(*i, j*)**: Pairs of variables for comparing regression coefficients.

**Output:**

Statistical measures for effect size differences including mean, standard error, confidence intervals, and *P*-values.

**Steps**

1. **Initialize Storage:**

   Create empty lists to store coefficients ***bp(i)*** and ***bp(j)***.

2. **Perform Bootstrapping:**

   **For k = 1 to n_*bootstrap* do**:

   - **Sample With Replacement:** Generate a bootstrap sample $D_k$ from **Dataset D**.
   - **Regression Analysis:** Compute regression on ***D_k*** to derive coefficients.
   - **Store Coefficients:** Save coefficients for variables ***i*** and ***j*** into $b_p(i)_k$ and $b_p(j)_k$.

3. **Initialize Difference Lists:**

   Prepare empty lists for absolute differences $d_s$ and directional differences $d_d$.

4. **Calculate Differences:**

   **For k = 1 to n_bootstrap do**:

   Calculate absolute difference: $d_s_k = |bp(i)_k| - |bp(j)_k|$

   Calculate directional difference: $d_d_k = bp(i)_k - bp(j)_k$

   Append ***ds_k*** *to* ***ds*** and $d_d_k$ to ***dd***.

5. **Statistical Analysis:**

   **Mean:** Compute the average of $d_s$ and $d_d$.

   **Standard Error:** Determine the standard deviation of $d_s$ and $d_d$.

**95% Confidence Interval:** Estimate the 2.5th and 97.5th percentiles of $\boldsymbol{d_s}$ and $\boldsymbol{d_d}$.

**Significance Test:**

Calculate proportions of $\boldsymbol{d_s}$ and $\boldsymbol{d_d}$ values $\leqslant 0$ and $\geqslant 0$.

Determine the minimum of these proportions.

Multiply the minimum proportion by 2 for the *P*-value.

6. **Output Results:**

   Present the computed mean, standard error, confidence intervals, and *P*-values for both $\boldsymbol{d_s}$ and $\boldsymbol{d_d}$.

We executed the program on the HINTS 5, Cycle 4 dataset within an experimental environment consisting of Python 3.11 and PyTorch 2.12. The computations were carried out on a server equipped with Intel Xeon Gold 6138 processors and NVIDIA Tesla V100 GPUs. The experimental results are presented in Tables 3, 4, and 5.